\documentclass[showpacs,aps,superscriptaddress,twocolumn]{revtex4}
\usepackage{mathrsfs}
\usepackage{array}
\usepackage{amsmath}
\usepackage{graphicx}
\usepackage{subfigure}
\usepackage{amstext}
\usepackage{amsfonts}
\usepackage{amssymb}
\usepackage{bm}
\usepackage{epstopdf}

\usepackage[usenames,dvipsnames]{color}
\usepackage{hyperref}
\usepackage{calc}
\newcommand{\C}[1]{\mathcal{#1}}
\newcommand{\+}[0]{\dagger}
\newcommand{\T}[0]{\intercal}
\newcommand{\x}[0]{{\cdot}}
\newcommand{\e}[0]{\bm{\epsilon}}
\newcommand{\dd}[0]{\mathrm{d}}
\newcommand{\sys}[0]{\textsc{s}}
\newcommand{\env}[0]{\textsc{e}}
\newcommand{\tot}[0]{\mathrm{tot}}
\newcommand{\Tr}[0]{\operatorname{Tr}}
\newcommand{\bra}[1]{\langle #1 |}
\newcommand{\ket}[1]{| #1 \rangle}
\newcommand{\braket}[1]{\langle #1 \rangle}
\newcommand{\mat}[1]{\begin{pmatrix} #1 \end{pmatrix}}
\newcommand{\mmat}[1]{\left(\begin{smallmatrix} #1 \end{smallmatrix}\right)}

\begin{document}
\title{Dissipative topological systems } 
\author{Yu-Wei Huang}
\affiliation{Department of Physics, 
National Cheng Kung University, Tainan 70101, Taiwan}
\author{Pei-Yun Yang}
\affiliation{Department of Chemistry, Massachusetts Institute of Technology, Cambridge, MA 02139, USA}
\author{I-Chi Chen}
\affiliation{Department of Physics and Astronomy, Iowa State University, Ames, Iowa 50011, USA}
\author{Wei-Min Zhang}
\email{wzhang@mail.ncku.edu.tw}
\affiliation{Department of Physics, 
National Cheng Kung University, Tainan 70101, Taiwan}

\begin{abstract}
Topological phases of matter are protected from local perturbations and therefore have been thought to be robust against decoherence.  
However, it has not been systematically explored whether and how topological states are dynamically robust against the environment-induced  decoherence.  
In this Letter, we develop a theory for topological systems that incorporate dissipations, noises and thermal effects.
We derive novelly the exact master equation and the transient quantum transport for the study of dissipative topological systems, 
mainly focusing on noninteracting topological insulators and topological superconductors.  The resulting exact master equation 
and the transient transport current are also applicable for the systems initially entangled with environments.
We apply the theory to the topological Haldane model (Chern insulator) and the quantized Majorana conductance to explore 
topological phases of matter that incorporate dissipations, noises and thermal effects, and demonstrate 
the dissipative dynamics of topological states.
\end{abstract}

\maketitle

Topological phases of matter are the most active research fields in modern condensed matter physics today \cite{KT1972,TKNN1982,Haldane1988,Wen1991,Kitaev2001,Kane2005,Bernevig2006,Bernevig2013,Wen2017,Chiu2016}.  
They comprise several exotic quantum phases such as topological insulators  and superconductors \cite{Hasan2010,Qi2011}, 
Weyl semimetals \cite{Xu2015},  fractional quantum Hall effect \cite{Tsui1982} and Majorana zero modes \cite{HZhang2018}, etc.  
These quantum phases of matter have been largely explored during the last decade 
\cite{Bernevig2013,Wen2017,Chiu2016,Hasan2010,Qi2011,Ando2015,Lutchyn2018}.  
However, realistic systems in nature have inevitable interactions with the surrounding environments.  
When system-environment interactions are not negligible, the dynamics of the systems are strongly 
influenced by dissipations and noises, which has become the main obstacle in practical realizations 
of quantum computing.  Although topological phases of matter have been thought to be robust 
against the environment-induced decoherence, a theory that incorporates dissipations, noises and thermal 
effects for demonstrating such robustness has been barely established.  In this Letter, we attempt to develop 
a dissipative quantum theory for topological phases of matter.

In contrast to an isolated quantum system, whose states are governed by Schr\"{o}dinger equation, the quantum 
state evolution of an open quantum system (the system interacting with environment) is determined by the master equation \cite{Breuer08,Weiss08}.  
Exact master equation for arbitrary open systems has only been formally formulated through the operator-projection 
method by Nakajima \cite{Nakajima58} and Zwanzig \cite{Zwanzig60}.  However, in practice, very few systems 
can be solved from Nakajima-Zwanzig master equation \cite{Breuer08,Weiss08}.  Therefore, most of investigations 
for open quantum system dynamics are often based on the Born-Markov approximation with Lindblad-type master 
equation \cite{Lindblad76,GKS76}, including some recent applications to topological systems \cite{Viyuela2012,Cheng2012,Rivas2013}.  
These investigations are valid only in the weak system-environment coupling regime.

There are some exceptions that one can derive the exact master equation for open quantum systems, using Feynman-Vernon 
influence functional approach \cite{Feynman63}.  A prototype example is the quantum Brownian motion (QBM), 
its exact master equation has been derived \cite{Leggett83,Haake85,HPZ1992,Grabert88} in 1980's-1990's. 
In the last decade, the exact master equation has also been derived for a large class of open systems described 
by particle-particle exchanges between the system and environments 
for both boson and fermion open systems \cite{Tu2008,Jin2010,Lei2012,Zhang2012,Zhang2018}, from which we 
also obtain the transient quantum transport theory that can reproduce explicitly the Schwinger-Keldysh's non-equilibrium Green 
function technique \cite{Haug2008,Yang2017}.  Very recently, we have extended the exact master equation for 
Majorano zero modes influenced by the gate-induced charge fluctuations \cite{Lai2018,Schmidt2012}.

In this Letter, we will derive novelly the exact master equation and the transient quantum transport for 
noninteracting topological insulators 
incorporating with initial system-reservoir entanglement.~Then we generalize the theory to topological 
superconductors with Bogoliubov-de~Gennes Hamiltonian that has potential applications in topological 
quantum computing. As a result, a dissipative quantum theory for topological phases of matter is established.  
We take the topological Haldane model (Chen insulator) \cite{Haldane1988,Jotzu2014} 
and the quantized Majorana conductance in superconductor-semiconductor hybrid systems \cite{HZhang2018,Lutchyn2018} 
as two type applications, to clarify the role of dissipations, noises and  thermal effects in topological phases of matter.

\vspace{0.2cm}
\noindent
\textit{1.~Open quantum systems with initial system-environment entanglement for noninteracting topological insulators.}
We begin with the open systems (either bosons or fermions) coupled to their environments that are described 
by the following Hamiltonian,
\begin{align}
H_\tot(t)
& = H_\sys(t) + H_\env(t) + H_{\sys\env}(t) \notag\\
& = \mat{\bm{a}^\+ & \bm{c}^\+} \x
    \mat{\e_\sys(t) & \bm{V}_{\sys\env}(t) \\ \bm{V}^\+_{\sys\env}(t) & \e_\env(t)} \x
    \mat{\bm{a} \\ \bm{c}}, \label{FAH}
\end{align}
where $H_\sys(t)$ and $H_\env(t)$ are the Hamiltonian of the system and the environment, respectively, 
and  $H_{\sys\env}(t)$ is the interaction between them.
The notation $\bm{a} \equiv (a_1, a_2, a_3, \dotsc)^\T$ is a one-column matrix and $a_i$ is the annihilation 
operator of the $i$-th energy level of the system.  Similarly, $\bm{c} \equiv (c_{k}, c_{k'}, c_{k''}, \dotsc)^\T$ 
and $c_k$ is the annihilation operator of the continuous spectrum mode $k$ of the environment, 
while $\e_\sys(t)$ and $\e_\env(t)$ are the spectra of the system and the environment, respectively, 
and $\bm{V}_{\sys\env}(t)$ is the coupling strength matrix between them.

Equation (\ref{FAH}) is applicable to both topological and non-topological open quantum systems. 
Topological structures can be manifested through energy eigen-wavefunctions. For open  systems, 
states of the system are described by the reduced density matrix which is determined from the total density 
matrix (a highly entangled state) of the system and environment: $\rho_\sys(t) \equiv \Tr_\env[\rho_\tot(t)]$.  
The total density matrix is governed by the von~Neumann equation: 
$\frac{\dd}{\dd t} \rho_\tot(t) = \frac{1}{i\hbar} [H_\tot(t), \rho_\tot(t)]$.
Taking a partial trace over the environment states from the von~Neumann equation, we have
\begin{align}
\frac{\dd}{\dd t} \rho_\sys(t)
=   \frac{1}{i\hbar} & [H_\sys(t), \rho_\sys(t)]
    + \bm{a}^\+ \x \bm{A}(t) + \bm{A}^\+(t) \x \bm{a} \notag\\
    & - ( \bm{a} \x \bm{A}^\+(t) + \bm{A}(t) \x \bm{a}^\+ )
, \label{fsms}
\end{align}
where the collective operator $\bm{A}(t) \equiv \frac{1}{i\hbar} \Tr_\env[\bm{V}_{\sys\env}(t) \x \bm{c} \rho_\tot(t)]$ 
which contains all the influence of the environment on the system dynamics. Here we have also used the fact 
that $\Tr_\env[H_\env(t), \rho_\tot(t)] = 0$.  Our aim is to carry out explicitly the partial trace in the collective 
operator $\bm{A}(t)$, from which the master equation can be novelly and straightforwardly obtained, 
and also the noise, thermal effects and dissipations in topological phases of matter can be explicitly explored.

For the initial system-environment decoupled or partitioned states \cite{Leggett1987}: $\rho_\tot(t_0) =\rho_\sys(t_0) 
\otimes \rho_\env(t_0)$, where $\rho_\env(t_0) =\frac{1}{Z} e^{-\beta H_\env(t_0)}$ is the thermal state of the environment, 
the exact master equation of Eq.~(\ref{fsms}) has been derived \cite{Tu2008,Jin2010,Lei2012,Zhang2012}, 
and the partial trace in the operator $\bm{A}(t)$ has also been explicitly computed \cite{Jin2010} using the 
Feynman-Vernon influence functional \cite{Feynman63}.
Now we consider the system and the environment in a partition-free initial state, $\rho_\tot(t_0) = \frac{1}{Z} 
e^{-\beta H_\tot(t_0)}$.  In this situation, the system is highly entangled with the environment from the 
beginning so that the Feynman-Vernon influence functional \cite{Feynman63} is no longer applicable.  
In experiments, most of realistic open quantum systems start with a partition-free initial state.  
Typical examples are various quantum devices which are usually equilibrated to the environment before 
one starts to manipulate them.  One often uses different quench methods to drive the system away from 
the equilibrium state to control the states of the system or to study its nonequilibrium dynamics.
This can be practically realized by the time-dependent parameters in Eq.~(\ref{FAH}).

Because of the quadratic nature of Eq.~(\ref{FAH}), with the explicit time-dependent Hamiltonian $H_\tot(t)$, the 
total density matrix is drived away from the initially entangled equilibrium state $\rho_\tot(t_0) = \frac{1}{Z} 
e^{-\beta H_\tot(t_0)}$, but it always lives in a Gaussian-type state.  Therefore, in coherent state representation \cite{zhang90}, 
we have
$
\bra{\bm{\xi}} \rho_\tot(t) \ket{\bm{\xi}'} = \frac{1}{Z(t)} \exp(\bm{\xi}^\+ \x \bm{\Omega}(t) \x \bm{\xi}')
$,
where $\ket{\bm{\xi}} \equiv \exp(\bm{a}^\+ \x \bm{\xi}_\sys + \bm{c}^\+ \x \bm{\xi}_\env) \ket{0}$ is the unnormalized 
coherent eigenstates of the particle annihilation operators $(\bm{a}, \bm{c})$ with eigenvalue $\bm{\xi} = (\bm{\xi}_\sys, 
\bm{\xi}_\env)$ which are complex numbers for bosons and Grassmann numbers for fermions, and
$
\bm{\Omega}(t) = \mmat{\bm{\Omega}_{\sys\sys}(t) & \bm{\Omega}_{\sys\env}(t) \\ \bm{\Omega}_{\env\sys}(t) & \bm{\Omega}_{\env\env}(t)}
$
is the Gaussian kernel of the total density matrix. 
By partially tracing over all the environment states, it is not difficult to find that
$
\bra{\bm{\xi}_\sys} \bm{A}(t) \ket{\bm{\xi}_\sys'}
=   \frac{1}{i\hbar} \bm{V}_{\sys\env}(t) \x (\bm{1} \mp \bm{\Omega}_{\env\env}(t))^{-1}
    \x \bm{\Omega}_{\env\sys}(t) \x \bm{\xi}_\sys' \bra{\bm{\xi}_\sys} \rho_\sys(t) \ket{\bm{\xi}_\sys'}
$,
from which we obtain:
\begin{equation}
\bm{A}(t)
=   \frac{1}{i\hbar} \bm{V}_{\sys\env}(t) \x (\bm{1} \mp \bm {\Omega}_{\env\env}(t))^{-1}
    \x \bm{\Omega}_{\env\sys}(t) \x \rho_\sys(t) \bm{a}
, \label{cas1}
\end{equation}   
where the upper (lower) sign of $\mp$ correspond to boson (fermion) systems.  Substituting this result into 
Eq.~(\ref{fsms}), we novelly obtain the exact master equation for the reduced density matrix of the system.

However, the key ingredient in the derivation of the master equation is to characterize explicitly the dissipation 
and noises induced by the environment, which are embedded in the time-dependent Gaussian kernel  
$\bm{\Omega}(t)$. Our aim is to find the relation between $\bm{\Omega}(t)$ and the physical measurable 
quantities such that dissipation and noise dynamics can be observed.
Note that under the Gaussian state, the Wick's theorem is always valid, and higher-order correlation functions 
can always be decomposed in terms of the single-particle correlations. A direct calculation shows that 
\begin{subequations}
\begin{align}
& \bm{n}_\sys(t)
=   \bm{\Omega}_\sys(t) \x (\bm{1}\mp\bm{\Omega}_\sys(t))^{-1}, \\
& \bm{n}_{\env\sys}(t)
=   (\bm{1} \mp \bm{\Omega}_{\env\env}(t))^{-1} \x \bm{\Omega}_{\env\sys}(t) \x (\bm{1} \mp \bm{\Omega}_\sys(t))^{-1}.
\end{align}
\end{subequations}
where $n_{\sys,ij}(t) \equiv \braket{a_j^\+(t) a_i(t)}$ and $n_{\env\sys,ki}(t) \equiv \braket{a_i^\+(t) c_k(t)}$ are the single particle correlations, and
$
\bm{\Omega}_\sys(t)
=   \bm{\Omega}_{\sys\sys}(t) \pm \bm{\Omega}_{\sys\env}(t) \x (\bm{1} \mp \bm{\Omega}_{\env\env}(t))^{-1} \x \bm{\Omega}_{\env\sys}(t)
$
which is given by
$
\bra{\bm{\xi}_\sys} \rho_\sys(t) \ket{\bm{\xi}_\sys'}
=   \frac{1}{Z_\sys(t)} \exp( \bm{\xi}_\sys^\+ \x \bm{\Omega}_\sys(t) \x \bm{\xi}_\sys' )$, 
and from which one can also prove that $ \bm{a} \rho_\sys(t) = \rho_\sys(t) \bm{\Omega}_\sys(t) \x \bm{a}$.

Furthermore, the time evolution of the system operators $a_i(t)$ can be directly solved from 
Eq.~(\ref{FAH}) with the Heisenberg equation of motion.  The solution can be written as $\bm{a}(t) = \bm{u}(t,t_0) \x \bm{a}(t_0) + \bm{F}(t)$, 
where $u_{ij}(t,t_0) \equiv \braket{[a_i(t), a_j^\dag(t_0)]_{\mp}}$ is the retarded Green function that 
describes the dissipation, and $F_i(t)$ linearly depends on $c_k(t_0)$ that characterizes noises,
see the explicit solution given in supplemental materials \cite{SM}.
Then 
\begin{equation}
\bm{n}_\sys(t)
=   \bm{u}(t,t_0) \x \bm{n}_\sys(t_0) \x \bm{u}^\+(t,t_0) + \bm{v}(t,t),
\end{equation}
where $\bm{v}(t,t)$ generalizes the Keldysh's correlation Green function that also includes initial 
system-environment entanglement \cite{Yang2015}.
Also, the electron transient current flowing from the environment into the system is
\begin{align}
I(t)
& \equiv -e \frac{\dd}{\dd t} \braket{\bm{c}^\+(t) \x \bm{c}(t)}
=   \frac{e}{i\hbar} \bm{V}_{\sys\env}(t) \x \bm{n}_{\env\sys}(t) + \text{H.c.} \notag\\
& = -e [ \bm{\kappa}(t,t_0) \x \bm{n}_\sys(t) + \bm{\lambda}(t,t_0) + \text{H.c.} ]
, \label{tc}
\end{align}
where the dissipation and noise coefficients $\bm{\kappa}(t,t_0) = \frac{1}{i\hbar} \e_\sys(t) - 
\dot{\bm{u}}(t,t_0) \x \bm{u}^{-1}(t,t_0)$ and $\bm{\lambda}(t,t_0) = \dot{\bm{u}}(t,t_0) \x \bm{u}^{-1}(t,t_0) 
\x \bm{v}(t,t) - \dot{\bm{v}}(t,t)$ are also determined explicitly by Green functions $\bm{u}(t,t_0)$ and $\bm{v}(t,t)$.
Combining all the above results together, Eq~(\ref{cas1}) becomes
\begin{equation}
\bm{A}(t)
= \! - [\bm{\kappa}(t,t_0) \x \bm{a} \rho_\sys(t) \!+\! \bm{\lambda}(t,t_0) \x (\rho_\sys(t) \bm{a} \mp \bm{a} \rho_\sys(t))]
, \label{as}
\end{equation}
which captures explicitly all the dissipation and noises induced by the environment.
The master equation (\ref{fsms}) and the transient current (\ref{tc}) simply become
\begin{subequations}
\label{current_Mas} 
\begin{align}
& \frac{\dd\rho_\sys(t)}{\dd t}
=   \frac{1}{i\hbar} \big[ H_\sys(t), \rho_\sys(t) \big]
    + \big[ \C{L}^{+}(t) + \C{L}^{-}(t) \big] \rho_\sys(t)
, \label{eme} \\
& I(t)
=   e \Tr_\sys \big[ \C{L}^{+}(t) \rho_\sys(t) \big]
=   -e \Tr_\sys \big[ \C{L}^{-}(t) \rho_\sys(t) \big],
\end{align}
\end{subequations}
where the current superoperators $\C{L}^{+}(t) \rho_\sys(t) = \bm{a}^\+ \x \bm{A}(t) + \bm{A}^\+(t) \x \bm{a}$ and $\C{L}^{-}(t) \rho_\sys(t) 
= -  \bm{a} \x \bm{A}^\+(t) - \bm{A}(t) \x \bm{a}^\+ $ carry the information current flowing into and out of the system, respectively.

It is easy to check that for fermionic systems, Eq.~(\ref{current_Mas}) reproduces the exact master equation and the transient transport 
current incorporating with the initial system-environment correlations given in Ref.~\cite{Yang2015}; For noninteracting bosonic systems, 
except for a special case \cite{Tan2011}, this gives a general dissipative 
theory incorporating initial system-environment entanglement.  The master equation and the transient current also have the same 
universal form derived from the Feynman-Vernon influence functional for the case of no initial system-environment entanglement 
\cite{Tu2008,Jin2010,Lei2012,Zhang2012}, whereas the initial system-environment entanglement is fully embedded into the correlation 
Green function $\bm{v}(t,t)$, as shown in \cite{Tan2011,Yang2015}.

\vspace{0.2cm}
\noindent
\textit{2.~Open systems for topological superconductors.}
Now, we generalize the exact master equation to the open systems containing paring couplings to the environment, such as the 
superconductor-semiconductor hybrid systems in the study of topological quantum computing.  Through a Bogoliubov transformation, 
the paring terms in the Hamiltonian of the system or the environment can always be switched into the coupling Hamiltonian 
between the system and the environment. Then the general Hamiltonian can be expressed as
\begin{align}
H_\tot(t)
=  & \bm{a}^\+ \x \e_\sys(t) \x \bm{a} + \bm{c}^\+ \x \e_\env(t) \x \bm{c} \notag \\
    & + \mat{\bm{a}^\+ & \bm{a}}  \mat{\bm{V}_{\sys\env}(t) & \bm{\Delta}_{\sys\env}(t) \\ \pm \bm{\Delta}_{\sys\env}^*(t) & \pm \bm{V}_{\sys\env}^*(t)}\mat{\bm{c} \\ \bm{c}^\+},
\end{align}
where the last term
is the Bogoliubov-de~Gennes Hamiltonian matrix describing effectively the pairing processes between the system and environment.

Following the same procedure, taking a partial trace over the environmental states from the von Neumann equation,
one obtains the same master equation (\ref{fsms}) for the reduced density matrix $\rho_\sys(t)$.  The only difference is the collective operator $\bm{A}(t)$ which is now given by
\begin{align}
\bm{A}(t)
\equiv \frac{1}{i\hbar} \Tr_\env[(\bm{V}_{\sys\env}(t) \x \bm{c} + \bm{\Delta}_{\sys\env}(t) \x \bm{c}^\+) \rho_\tot(t)].
\end{align}
Similarly, if one can carry out explicitly the partial trace over the environmental states for the above operator $\bm{A}(t)$, the exact 
master equation involving pairing couplings can also be novelly and straightforwardly obtained. Indeed, using the same procedure, we obtain
\begin{align}
\mat{\bm{A}(t) \\ -\bm{A}^\+(t)}
=   & - \C{K}(t,t_0) \mat{\bm{a} \rho_\sys(t) \\ \bm{a}^\+ \rho_\sys(t)} \notag\\
    & - \C{Z} \Lambda(t,t_0) \mat{
        \rho_\sys(t) \bm{a} \mp \bm{a} \rho_\sys(t) \\
        \bm{a}^\+ \rho_\sys(t) \mp \rho_\sys(t) \bm{a}^\+
    }
, \label{tsa}
\end{align}
where $\C{Z} = \mmat{\bm{1} & \bm{0} \\ \bm{0} & \mp\bm{1}}$, and $\C{K}(t,t_0)$ and  $\Lambda(t,t_0)$ are given later, see Eq.~(\ref{sopc}).  
Thus, the master equation for topological superconductor open systems with arbitrary pairing couplings has exactly the same form as 
Eq.~(\ref{eme}) but the collective operator $\bm{A}(t)$ is modified by Eq.~(\ref{tsa}).

Because the topological superconductor open systems involving pairing interactions, the explicit form of the master equation is more complicated.
Substituting the solution of Eq.~(\ref{tsa}) into Eq.~(\ref{fsms}), we get
\begin{align}
\frac{\dd}{\dd t} \rho_\sys(t) 
=  &  \frac{1}{i\hbar} [H'_{\sys}(t,t_0), \rho_\sys(t)] \notag\\
    & + \sum_{ij} \gamma_{ij}(t,t_0) \begin{aligned}[t] \big[
        & 2 a_j \rho_\sys(t) a_i^\+ \\
        & - \rho_\sys(t) a_i^\+ a_j - a_i^\+ a_j \rho_\sys(t)
    \big] \end{aligned} \notag\\
    & + \sum_{ij} \widetilde{\gamma}_{ij}(t,t_0) \begin{aligned}[t] \big[
        & a_i^\+ \rho_\sys(t) a_j \pm a_j \rho_\sys(t) a_i^\+ \\
        & - \rho_\sys(t) a_j a_i^\+ \mp a_i^\+ a_j \rho_\sys(t)
    \big] \end{aligned} \notag\\
    & \mp \frac{1}{2} \sum_{ij} \big\{ \lambda_{ji}(t, \begin{aligned}[t]
        &t_0) \big[ 2 a_j^\+ \rho_\sys(t) a_i^\+ - a_i^\+ a_j^\+ \rho_\sys(t) \\
        & - \rho_\sys(t) a_i^\+ a_j^\+ \big] + \text{H.c.} \big\} \end{aligned}
    \label{pairMEM}
\end{align}
The first term is the renormalized Bogoliubov-de~Gennes Hamiltonian of the system
$
H_\sys'(t,t_0)
=   \frac{1}{2} \mmat{\bm{a}^\+ & \bm{a}}
    \mmat{\bm{E}_\sys'(t,t_0) & \bm{\Delta}_\sys'(t,t_0) \\ \bm{\Delta}_\sys'^\+(t,t_0) & \pm \bm{E}_\sys'^\T(t,t_0)}
    \mmat{\bm{a} \\ \bm{a}^\+}
$.
The second and the third terms describe the dissipation and noise dynamics which are very similar to 
the cases without including pairings \cite{Tu2008,Jin2010,Lei2012,Yang2015}.  The last term comes from pairing-process 
induced dissipation.  Explicitly, those time non-local dissipation and noise coefficients 
\begin{subequations}
\label{sopc}
\begin{align}
& \mat{\bm{E}_\sys'(t,t_0) & \bm{\Delta}_\sys'(t,t_0)  \\ \bm{\Delta}_\sys'^\+(t,t_0) & \pm \bm{E}_\sys'^\T(t,t_0)}
=   \C{E}_\sys(t)   + \frac{\hbar}{2i} ( \C{K}(t) - \C{K}^\+(t) )\notag\\
& \quad\quad
=   - \frac{\hbar}{2i} ( \C{Z} \dot{\C{U}}(t,t_0) \C{U}(t,t_0)^{-1} - \text{H.c.} ), \\
& \mat{\bm{\gamma}(t,t_0) & \bm{\gamma}'(t,t_0) \\ \bm{\gamma}'^\+(t,t_0) & \mp \bm{\gamma}^\T(t,t_0)}
=   \frac{1}{2} ( \C{K}(t,t_0) + \C{K}^\+(t,t_0) ) \notag\\
& \quad\quad
=   - \frac{1}{2} ( \C{Z} \dot{\C{U}}(t,t_0) \C{U}(t,t_0)^{-1}+ \text{H.c.} ), \\
& \mat{
    \widetilde{\bm{\gamma}}(t,t_0) & \bm{\lambda}(t,t_0) \\
    \bm{\lambda}^\+(t,t_0) & 2\bm{\gamma}^\T(t,t_0) {\pm} \widetilde{\bm{\gamma}}^\T(t,t_0)
}
=   - \Lambda(t,t_0) - \Lambda^\+(t,t_0) \notag\\
& \quad\quad
=   \tfrac{\dd}{\dd t} \C{V}(t,t) - ( \dot{\C{U}}(t,t_0) \C{U}(t,t_0)^{-1} \C{V}(t,t) + \text{H.c.} ),
\end{align}
\end{subequations}
are all determined by the retarded and correlation Green functions $\C{U}(t,t_0)$ and $\C{V}(t,t)$ incorporating pairing interactions \cite{SM}.
Those matrices are Hermitian, so we have $\bm{E}_\sys'^\+(t,t_0) = \bm{E}_\sys'(t,t_0)$ and $\bm{\Delta}_\sys'^\T(t,t_0) 
= \pm \bm{\Delta}_\sys'(t,t_0)$, $\bm{\gamma}^\+(t,t_0) = \bm{\gamma}(t,t_0)$ and $\bm{\gamma}'^\T(t,t_0) = \mp \bm{\gamma}'(t,t_0)$,
$\widetilde{\bm{\gamma}}^\+(t,t_0) = \widetilde{\bm{\gamma}}(t,t_0)$ and $\bm{\lambda}^\T(t,t_0) = 2\bm{\gamma}'(t,t_0) \pm \bm{\lambda}(t,t_0)$.
The experimentally measured transport current flowing from the environment into the system is given by
\begin{align}
I(t) = \frac{e}{i\hbar} \bm{V}_{\sys\env}(t) \x \bm{n}_{\env\sys}(t) - \frac{e}{i\hbar} \bm{\Delta}_{\sys\env}(t) \x \bm{q}_{\env\sys}(t) + \text{H.c.}
, \label{pairingcurrent}
\end{align}
where $q_{\env\sys,ij}(t) \equiv \braket{a_j^\+(t) c_i^\+(t)}$.
From this transient current one can study Majorana quantum transport dynamics that we will discuss latter.

\begin{figure}
\centering
\subfigure[Haldane model]{\includegraphics[width=0.3\linewidth]{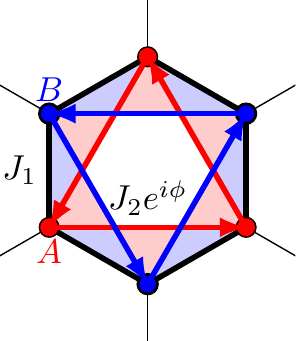}}
\subfigure[Phase diagram]{ \includegraphics[width=0.4\linewidth]{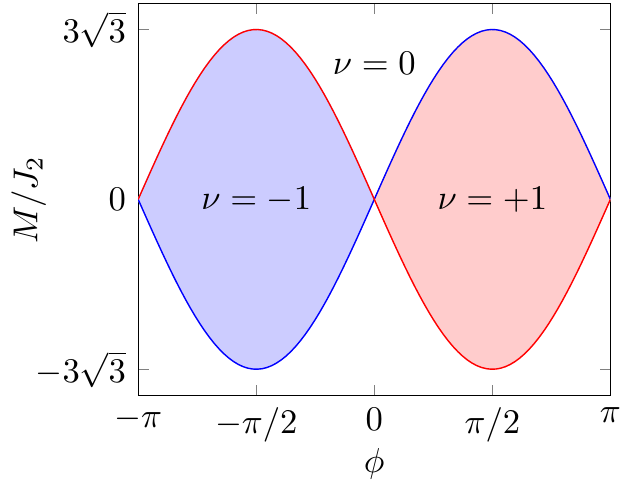}}
\subfigure[Adatom at bulk/zigzag edge state]{\includegraphics{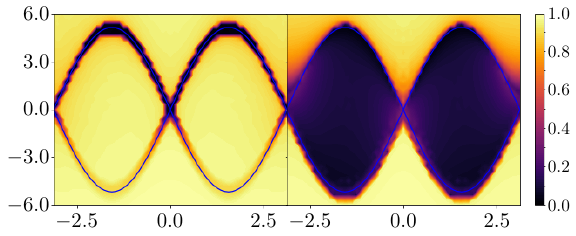}}
\caption{\label{pdhm}
(a) Haldane model with honeycomb lattices.
(b) Topological phase diagram of the Haldane model with Chern number $\nu$ by varying the parameters $M$ and $\phi$ in Eq.~(\ref{hmh}). 
(c) The steady-state values of the occupation number of the adatom in the bulk state (left) and in the zigzag edge state (right), 
respectively, solved from the master equation.}
\end{figure}

\vspace{0.2cm}
\noindent
\textit{3.~Applications.}
The first application is the topological Haldane model which describes quantum Hall effect in honeycomb lattice without 
magnetic field \cite{Haldane1988} and has been experimentally realized with ultracold fermionic atoms \cite{Jotzu2014}, 
and its Hamiltonian can be written as
\begin{align}
H
= & M \sum_{i} \left( a_{i}^\+ a_{i} - b_{i}^\+ b_{i} \right)
    - J_1 \sum_{\langle i,j \rangle} \left( a_{i}^\+ b_{j} + \text{H.c.} \right) \notag\\
    & + J_2 \sum_{\langle\!\langle i,j \rangle\!\rangle} \left(
        e^{i\phi} a_{i}^\+ a_{j} + e^{-i\phi} b_{i}^\+ b_{j} + \text{H.c.}
    \right)
, \label{hmh}
\end{align}
where $a_i$ ($b_i$) is the annihilation operator of A (B) site electrons, and $J_1$ and $J_2$ are the nearest neighbor 
and the next-nearest neighbor coupling strengths, respectively, see Fig.~\ref{pdhm}(a).
The energy difference $M$ between A and B sites breaks inversion symmetry, and the phase $\phi$ in the next-nearest 
neighbor couplings breaks time-reversal symmetry that topologically leads to quantum Hall effect. The  
non-trivial topological phase is located in the region of $|M| < 3\sqrt{3}|J_2\sin\phi|$ (see Fig.~\ref{pdhm}(b)), in 
which the band gap is closed at the edge of lattice.  Here we attempt to dynamically probe this topological structure 
in Haldane model from the open quantum system point of view by coupling an adatom to the honeycomb lattices. 

Putting an adatom ($H_a = \epsilon_0 c_d^\+ c_d$) on the edges or bulk of lattices, described by the coupling Hamiltonian 
$H_I = V c_d^\+ a_j+ \text{H.c.}$, where $j$ is the coupled site, we can study the dissipative dynamics of the 
adatom under the influence of the topological structure of the Haldane model.
We treat the honeycomb lattice with the Haldane Hamiltonian (\ref{hmh}) as the environment of the adatom.  
Then the solution of the reduced density matrix of the adatom can be determined effectively by the occupation 
number $n_a(t) = \Tr_\sys[c_d^\+ c_d \rho_\sys(t)] $.  By dynamically solving the occupation number of the 
adatom (initially occupied), we find that its steady-state solution manifests the whole topological structure of the Haldane model, as shown 
in Fig.~\ref{pdhm}(c), as a result of dissipation. In Fig.~\ref{pdhm}(c) the dark color corresponds to the complete dissipation 
(zero occupation in the adatom in the steady-state limit but initially it is fully occupied) in the topological phase.
Such dissipation is built up only when the lattice energy gap closes, which occurs at the edge of non-trivial topological phase, see the right plot in Fig.~\ref{pdhm}(c).
This provides indeed a very useful method of probing topological structures for more complicated topological systems  
through the study of dissipative dynamics of  adatoms (impurities).

Another application is the quantized Majorana conductance in superconductor-semiconductor hybrid system that has 
been very recently observed \cite{HZhang2018}.  The Hamiltonian of the total system is modeled as a tight-binding 
$N$-site p-wave superconductor, with its left/right ends of superconductor coupled respectively with the left/right leads. 
One can solve the large number chain of superconductor with zero chemical potential \cite{Schmidt2012}, in which 
two Majorana zero modes are localized at the ends of the chain with exponentially decaying wave function along the chain.  Focusing 
only on the zero modes, we have the interaction Hamiltonian of the zero modes coupled with the two leads,
\begin{equation}
\begin{aligned}
H_I
& = \sum_k
    \mat{c_{L,k}^\+ & c_{L,k}}
    \mat{V'_{L,k} & \Delta'_{L,k} \\ -\Delta'_{L,k} & - V'_{L,k}}
    \mat{b_0 \\ b_0^\+} \\
& + \sum_k
    \mat{c_{R,k}^\+ & c_{R,k}}
    \mat{V'_{R,k} & -\Delta'_{R,k} \\ \Delta'_{R,k} & - V'_{R,k}}
    \mat{b_0 \\ b_0^\+}.
\end{aligned}
\end{equation}
where $b_0$ is zero mode annihilation operator, and $V'_{\alpha,k} = \frac{\sqrt{1-\delta^2}}{2} (1 + \delta^{(N-1)/2}) 
V_{\alpha,k}$, $\Delta'_{\alpha,k} = \frac{\sqrt{1-\delta^2}}{2} (1 - \delta^{(N-1)/2}) V_{\alpha,k}$.  It shows that the 
tunneling strength $V'_{\alpha,k}$ and pairing parameter $\Delta'_{\alpha,k}$ only depend on dimensionless parameter 
$\delta = (\Delta - w)/(\Delta + w)$.  And for large $N$, the tunneling strength is almost equal to the pairing parameter, 
which makes the superconducting system evolve into a half-filled state.

\begin{figure}
\centering
\includegraphics[width=\linewidth]{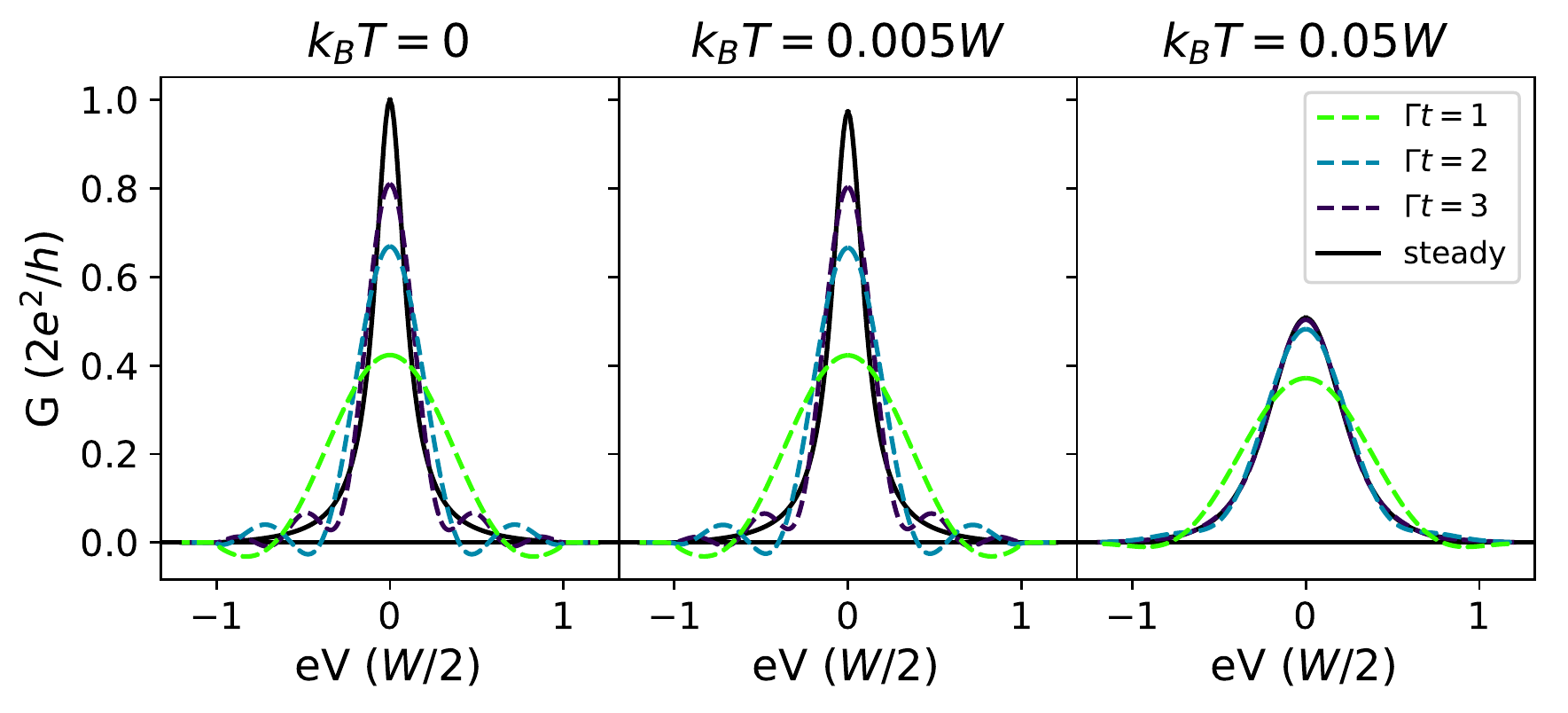}
\caption{\label{zbp}Buildups of zero-bias conductance peak in the time domain at (a) zero temperature, (b) low temperature ($k_BT = 0.005W$), 
and (c) high temperature ($k_BT = 0.05W$).  Here we take $J'(\epsilon) = \Gamma \sqrt{1 - (\frac{\epsilon}{W/2})^2}$, 
which is solved from the tight-binding model, and set height-width ratio as $\Gamma/W = 0.095$.
}
\end{figure}

Applying bias to the leads, one can measure the current through the superconductor.  
From Eq.~(\ref{pairingcurrent}), we study the transient current and transient differential conductance of superconductor,
and find a relation between the spectral density and the conductance in the steady state limit. 
Especially when two spectral densities $J_{\alpha}(\epsilon)$ are same and symmetric, we have,
\begin{align}
G(\mu,t\to\infty) = {\textstyle\int \dd\epsilon} \, G_0(\epsilon) \tfrac{\partial}{\partial\mu} f(\mu,\epsilon),
\end{align}
where $G_0(\mu) = 1/\big[\big(\tfrac{\mu - \Delta'(\mu)}{J'(\mu)/2}\big)^2 + 1\big]$ is the conductance at zero temperature 
in the unit of $2e^2/h$, and $f(\mu,\epsilon)$ is the Fermi-Dirac distribution, $J'(\epsilon) = \sum_{\alpha,k} 2\pi 
(|V_{\alpha,k}'|^2 + |\Delta_{\alpha,k}'|^2) \delta(\epsilon_{\alpha,k} - \epsilon)$ is the spectral density, and $\delta \mu'(\epsilon) 
= \C{P}\int \frac{\dd\epsilon'}{2\pi} \frac{J'(\epsilon')}{\epsilon - \epsilon'}$ is the corresponding energy shift, which is 
anti-symmetric because of the symmetric spectral density.
It shows that at zero temperature, the conductance at zero bias is precisely the quantized Majorana conductance, $2e^2/\hbar$ \cite{Law2009}, 
recently measured in experiment \cite{HZhang2018}. It has a Lorentzian function shape deformed by the energy shift and the decay 
rate, and thermal fluctuations broaden and lower down the zero-bias peak by convolution.
The buildups of zero-bias peak are shown in Fig.~\ref{zbp}.  The transient behavior of current in different bias involves 
different frequencies, which induces the oscillation of differential conductance.  It shows that Majorana conductance is 
indeed the observation of the dissipation and thermal fluctuations of the Majorana zero mode.

In conclusion, we  novelly derive a dissipation theory for noninteracting topological systems, which allows one
to investigate the dynamics of topological states incorporating dissipations, noises and thermal effects. 
By applying the theory to the Haldane model and the quantized Majorana conductance in a superconductor-semiconductor hybrid system, 
we demonstrate how dissipation and noises make topological structures observed in experiments. 
On the other hand, dissipation and noises are the sources of decoherence. Therefore, topological states cannot be 
immune from decoherence. 


\end{document}